\documentclass[aps,prd,amsmath,amssymb,preprintnumbers,preprint,nofootinbib,11pt]{article}
\pdfoutput=1
\usepackage{epsfig}
\usepackage{colortbl}
\usepackage{amsthm}
\usepackage{graphics}
\usepackage{graphicx}

\usepackage{graphicx}
\usepackage{pxfonts}
\usepackage{epstopdf}
\usepackage{floatpag}
 \usepackage{colortbl}
 \usepackage{youngtab}
\usepackage{slashed}
\usepackage{epsfig}
\definecolor{rossoCP3}{cmyk}{0,.88,.77,.40}

\newcommand{\ft}{{\cal F}_t}
\renewcommand{\(}{\left(}
\renewcommand{\)}{\right)}

\def\beq{\begin{equation}}
\def\eeq{\end{equation}}
\def\be{\begin{equation}}
\def\ee{\end{equation}}
\def\bea{\begin{eqnarray}}
\def\eea{\end{eqnarray}}
\def\ba{\begin{eqnarray}}
\def\ea{\end{eqnarray}}

\definecolor{red}{rgb}{1.00,0.00,0.00}

\catcode`\@=11
\def\lsim{\mathrel{\mathpalette\@versim<}}
\def\gsim{\mathrel{\mathpalette\@versim>}}
\def\@versim#1#2{\vcenter{\offinterlineskip
\ialign{$\m@th#1\hfil##\hfil$\crcr#2\crcr\sim\crcr } }}
\catcode`\@=12

\catcode`@=12
\begin{document}
\thispagestyle{empty}
\vspace{0.01in}
\begin{center}
{\LARGE \bf \color{rossoCP3} Composite Inflation confronts BICEP2 and PLANCK\\}
\vspace{0.5in}
{\Large Phongpichit Channuie$^1$\\}
\vspace{0.1in}
{\sl $^1$ Department of Physics, School of Science, Walailak University, \\Thasala District, Nakhon Si Thammarat, 80161, Thailand\\}
\vspace{0.1in}
{$^1$ Email: phongpichit.ch@wu.ac.th\\} 
\vspace{0.3in}
{\Large Khamphee Karwan$^2$\\}
\vspace{0.1in}
{\sl $^2$ The Institute for Fundamental Study, Naresuan University,
\\Phitsanulok 65000 and Thailand Center of Excellence in Physics, \\Ministry of Education,
Bangkok 10400, Thailand\\}
\vspace{0.1in}
{$^2$ Email: khampheek@nu.ac.th\\} 
\end{center}
\vspace{0.4in}
\begin{abstract}\
We examine observational constraints on single-field inflation in which the inflaton is a composite field stemming from a four-dimensional strongly interacting field theory. We confront the predictions with the Planck and very recent BICEP2 data. In the large non-minimal coupling regions, we discover for MCI model that the predictions lie well inside the joint $68\%$ CL for the
Planck data, but is in tension with the recent BICEP2 observations.
In the case of the GI model, the predictions satisfy the Planck results. However, this model can produce a large tensor-to-scalar ratio consistent with the recent BICEP2 observations
if the number of e-foldings is slightly smaller than the range commonly used. For a super Yang-Mills paradigm, we discover that the predictions satisfy the Planck data,
and surprisingly a large tensor-to-scalar ratio consistent with the BICEP2 results can also be produced
for an acceptable range of the number of e-foldings and of the confining scale. In the small non-minimal coupling regions, all of the models can satisfy the BICEP2 results.
However, the predictions of GI and SgbI models cannot satisfy the observational bound on the amplitude of the curvature perturbation launched by Planck,
and the techni-inflaton self-coupling in the MCI model is constrained to be extremely small. \\\\
{PACS: 98.80.Cq,\,98.80-k\\} 
\end{abstract}

\newpage
\tableofcontents

\section{ Introduction}

It was widely excepted that there was a period of accelerating expansion in the very early universe. Such period is traditionally known as inflation.
The inflationary paradigm \cite{Alex,KSa,KSa1,DKa,GUT} tends to solve important issues, e.g. the magnetic monopoles, the flatness, and the horizon problems, plagued the standard big bang theory and successfully describes the generation and evolution of the observed large-scale structures of the universe.
The inflationary scenario is formulated so far by the introduction of (elementary) scalar fields (called inflaton) with a nearly flat potential (see, e.g.~\cite{new,new1,chaotic,natural,natural1,Linde}). 

However, the theories featuring elementary scalar fields are unnatural meaning that quantum corrections generate unprotected quadratic divergences which must be fine-tuned away if the models must be true till the Planck energy scale. Therefore, it would be of great interest to imagine natural models underlying the cosmic inflation. In general, however, the inflaton need not be an elementary degree of freedom.
Recent investigations show that it is possible to construct models in which the inflaton emerges as a composite state of a four-dimensional strongly coupled theory \cite{Channuie:2011rq,Bezrukov:2011mv,Channuie:2012bv}. These types of models have already been stamped to be composite inflation. There were other models of super or holographic composite inflation \cite{Cvetic:1989eg,Thomas:1995dq,GarciaBellido:1997mq,Allahverdi:2006iq,Hamaguchi:2008uy,Evans:2010tf,Alberghi:2010sn,Alberghi:2009kk}.

Practically, all speculative ideas concerning physics of very early universe can be falsified by using the observation of the large scale structure. In particular, the temperature fluctuations observed in the Cosmic Microwave Background (CMB) is basically regarded as providing a clear window to probe the inflationary cosmology.
To testify all inflationary models, we need observables including: (i) the scalar spectral index $n_{s}$, (ii) the amplitude of the power spectrum for the curvature perturbations,
(iii) tensor-to-scalar ratio $r$, (iv) the non-gaussianity parameter $f_{\rm NL}$.

Yet, other relevant parameters are the running of scalar spectral index $\alpha\equiv dn_{s}/d \ln k$ and the spectral index for tensor perturbations $n_{\rm T}$.
Most recently, the Planck satellite data showed that the spectral index $n_{s}$ of curvature perturbations is constrained to be $n_{s}=0.9603\pm0.0073$ (68$\%$ CL) and ruled out the exact scale-invariance ($n_{s}=1$) at more than $5\sigma$ confident level (CL),
whilst the amplitude of the power spectrum for the curvature perturbations $|\zeta|^2$ is bounded to be ${\cal A}_{s} = 3.089^{+0.024}_{-0.027}$ (68$\%$ CL) \cite{Planck}\,with ${\cal A}_{s} \equiv \ln(|\zeta|^{2}\times 10^{10})$. Having constrained by Planck, the tensor-to-scalar ratio $r$ is bounded to be $r < 0.11$ (95\%CL). Surprisingly, the recent BICEP2 data renders the bound on $r$ to be $r=0.20^{+0.07}_{-0.05}$ with $r=0$ disfavored at $7.0\sigma$ CL \cite{Ade:2014xna}. The attempts to explain the seed perturbations from inflation were originally studied by many authors \cite{MuCh,SWHa,AAS1,GuPi}. As for gravitational waves generated during inflation, it was earlier calculated in \cite{AAS}.

The paper is organized as follows: In section (\ref{setup}), we first spell out the setup for a generic model of composite paradigm. We then derive equations of motion and useful expressions.
In section (\ref{constraint}), we derive $n_s$, $r$ and ${\cal A}_{s}$ for composite paradigms. In section (\ref{model}), we examine $n_s$, $r$ and ${\cal A}_{s}$, and the range of the model parameters in which these quantities satisfy the observational bound is evaluated. Using the Planck and very recent BICEP2 observations of the $(n_{s}-r)$ plane, we also confront our results with such observations. Finally, the conclusions are given in section (\ref{concl}).

\section{Composite Formulations and Background Evolutions}
\label{setup}

Recently, it has already been shown that cosmic inflation can be driven by four-dimensional strongly interacting theories non-minimally coupled to gravity \cite{Channuie:2011rq,Bezrukov:2011mv,Channuie:2012bv}.
The general  action for composite inflation in the Jordan frame takes the form for scalar-tensor theory of gravity as
\footnote{We used the signature of the matrix as $(-,+,+,+)$ throughout the paper.}
\begin{eqnarray}
\mathcal{S}_{\rm CI,J}=\int d^{4}x \sqrt{-g}\Big[\frac{M^{2}_{\rm P}}{2} F(\Phi) R - \frac{1}{2}G(\Phi)g^{\mu\nu} \partial_{\mu}\Phi\partial_{\nu}\Phi - V(\Phi)  \Big]. \label{action}
\end{eqnarray}
The functions $F(\Phi)$ and $G(\Phi)$ in this action are defined as
\begin{equation}
F(\Phi) = 1 + \frac\xi{M^{2}_{\rm P}}\,\Phi^{\frac{2}{D}} \quad {\rm and} \quad G(\Phi) = \frac{1}{D^2}G_{0}\Phi^\frac{2 - 2D}{D}\,,
\label{fg1}
\end{equation}
where $D$ is the mass dimension of the composite field $\Phi$, $G_{0}$ is a constant
and $1/D^2$ is introduced for later simplification.
In our setup, we write the potential in the following form:
\begin{equation}
V(\Phi) = \Phi^{4/D} f(\Phi)\quad\quad{\rm with}\quad\quad\Phi \equiv \varphi^{D} \,,
\label{fg}
\end{equation}
where the field $\varphi$ possesses unity canonical dimension and $f(\Phi)$ is a general function of the field $\Phi$ concretely implemented below. The non-minimal coupling to gravity is signified by the dimensionless coupling $\xi$.
Here, we write the general action for the composite inflation in the form of scalar-tensor theory of gravity in which the inflaton non-minimally couples to gravity. At the moment, the non-minimal term $\xi\Phi^{2/D}R/M^{2}_{\rm P}$ has purely phenomenological origin. The reason resides from the fact that one wants to relax the unacceptable large amplitude of primordial power spectrum generated if one takes $\xi=0$ or smaller. 

According to the above action, the Friedmann equation and the evolution equations for the background field are respectively given by
\begin{eqnarray}
3M^{2}_{\rm P}F H^{2} + 3M^{2}_{\rm P}\dot{F} H  =  3M^{2}_{\rm P}H^{2}F\left(1 + 2{\cal F}_{t}\right) = \frac{1}{2} G \dot{\Phi}^{2} + V(\Phi)\,, \label{h2}
\end{eqnarray}
\begin{eqnarray}
3M^{2}_{\rm P}F H^{2} + 2M^{2}_{\rm P}\dot{F} H + 2M^{2}_{\rm P}F \dot{H} + M^{2}_{\rm P}\ddot{F}=-\frac{1}{2} G \dot{\Phi}^{2} + V(\Phi)\,, \label{h21}
\end{eqnarray}
\begin{eqnarray}
G\ddot{\Phi} + 3HG\dot{\Phi} + \frac{1}{2} G_{\Phi}\dot{\Phi}^{2} + V_{\Phi}= 3M^{2}_{\rm P}F_{\Phi}\left(\dot{H} + 2H^{2}\right)\,,
\label{kg1}
\end{eqnarray}
where ${\cal F}_{t} = \dot{F}/(2HF)$, $H$ is the Hubble parameter, subscripts \lq\lq$\Phi$\rq\rq\,denote derivative with respect to $\Phi$, and the dot represents derivative with respect to time, $t$.
In the following calculations, we will set $M^{2}_{\rm P} = 1$.
In order to derive the observables, it is common to apply the standard slow-roll approximations such that 
\begin{eqnarray}
|\ddot{\Phi}/\dot{\Phi}|\ll H\,,\quad\quad|\dot{\Phi}/\Phi|\ll H\quad\quad{\rm and}\quad\quad|G\dot{\Phi}^{2}/2|\ll V(\Phi)\,.
\label{kg}
\end{eqnarray}
With these conditions, the equations given in Eq.(\ref{h2})-Eq.(\ref{kg}) become
\begin{eqnarray}
\epsilon = {\cal F}_{t} - \frac{V_{\Phi}}{V}\frac{F}{F_{\Phi}}{\cal F}_{t}\quad\quad{\rm and}\quad\quad\Phi'= \frac{1}{\left(1+ \frac{3F_{\Phi}^2}{2 F G}\right)}\Big(2 \frac{F_{\Phi}}{G} - \frac{V_{\Phi}}{V} \frac{F}{G}\Big)\,.
\label{epsi-ss}
\end{eqnarray}
Here the first relation can be derived by differentiating Eq.(\ref{h2})
with respect to time and then applying the above slow-roll conditions and $\Phi'\equiv d\Phi/d\ln a=\dot{\Phi}/H$. In order to keep our investigation more transparency, we will express hereafter the relevant parameters in term of the canonical-dimension one field $\varphi$. For more convenience, we expand the $\epsilon$ parameters around the coupling $\xi$, and find for $\xi\ll1$
\begin{eqnarray}
\epsilon \simeq \frac{(4+\Theta \varphi)^{2}}{2G_{0}\varphi^{2}}+\frac{1}{2G_{0}}\Big[-8+2\Theta\varphi+\Theta^{2}\varphi^{2}\Big]\xi+{\cal O}(\xi^2)\,, \label{ep-sxi}
\end{eqnarray}
and for $\xi\gg 1$
\begin{eqnarray}
\epsilon \simeq \frac{1}{12}\left[2\Theta \varphi+\Theta^{2}\varphi^{2}\right] + \frac{1}{72\xi\varphi^{2}}\Big[48+60\Theta \varphi +12\Theta^{2}\varphi^{2}-2G_{0}\Theta\varphi^{3}\nonumber\\\quad\quad\quad\quad\quad\quad\quad\quad\quad\quad\quad\quad\quad\quad\quad-G_{0}\Theta^{2}\varphi^{4}\Big]+{\cal O}(1/\xi^2)\,, \label{ep-lxi}
\end{eqnarray}
with $\Theta\equiv (1/f(\varphi)) (\partial f(\varphi) / \partial \varphi)$. We can also do the same exercise to expand $\varphi'$ around a small $\xi$ such that 
\begin{eqnarray}
\varphi' \simeq -\frac{1}{G_{0}\varphi}\Big[4+\Theta\varphi\Big] -\frac{\Theta\varphi^{2}}{G_{0}}\xi+{\cal O}(\xi^2)\,, \label{pp-sxi}
\end{eqnarray}
and for $\xi\gg 1$
\begin{eqnarray}
\varphi' \simeq -\frac{1}{6}\Theta\varphi^{2}+\frac{\varphi^{-1}}{36\xi}\Big[-24-12\Theta\varphi+G_{0}\Theta\varphi^{3}\Big]+{\cal O}(1/\xi^2)\,. \label{pp-lxi}
\end{eqnarray}
Having computed the field $\varphi$ at the end of inflation $\varphi_e$  by using the condition $\epsilon(\varphi_{e})=1$, one can determine the number of e-foldings via
\begin{eqnarray}
{\cal N}(\varphi) = \ln\frac{a_e}{a} = \int_{a}^{a_e}\frac 1{\tilde{a}}d{\tilde a} = \int_{t}^{t_e}H d{\tilde t} = \int_{\varphi}^{\varphi_e}\frac{H}{\dot{{\tilde \varphi}}}d{\tilde\varphi}=\int_{\varphi}^{\varphi_e}\frac 1{{\tilde \varphi}'}d{\tilde\varphi}\,,
\label{efold}
\end{eqnarray}
where the subscript \lq\lq $e$\rq\rq\,denotes the evaluation at the end of inflation
and $\varphi'$ is given by Eq.(\ref{epsi-ss}).
At the observable perturbation exits the horizon, we can evaluate the field $\varphi$ once the number of e-foldings ${\cal N}$ is specified.
Determining the value of $\varphi$ and $\varphi'$ when the perturbations exit the horizon allows us to compute the spectral index and power spectrum amplitude in terms of the number of e-foldings.

\section{Power Spectra and Spectral Index}
\label{constraint}
In order to obtain the power spectra for our models,
it is (tricky) convenient to use the results in Einstein frame. In order to transform the action in the Jordan frame into the Einstein one, we take the following transformation,
\be
\tilde{g}_{\mu\nu} = F\(\Phi\)g_{\mu\nu}\,.
\ee
Regarding to the above (conformal) implementation, the action in Eq.~(\ref{action}) can be written in Einstein frame as
\be
\mathcal{S}_{\rm CI,E}=\int d^{4}x \sqrt{-\tilde{g}}\Big[\frac{M^{2}_{\rm P}}{2} \tilde{R} - \frac{1}{2}\partial_{\mu}\tilde\Phi\partial^{\mu}\tilde\Phi - U(\tilde\Phi)  \Big], 
\label{action-e}
\ee
where $\tilde g$ and $\tilde R$ are computed from $\tilde{g}_{\mu\nu}$, \lq\lq tildes\rq\rq\,represent the quantities in the Einstein frame, and
\be
\frac{\partial \Phi}{\partial \tilde\Phi} = \frac F{\sqrt{G F + 3 F_{\Phi}^2 / 2}}\,,
\quad\mbox{and}\quad
U(\tilde\Phi) = \left. \frac{V(\Phi)}{F^2(\Phi)} \right|_{\Phi = \Phi(\tilde\Phi)}\,.
\ee
Using the expression for the slow-roll parameter, $\tilde\epsilon$, in the Einstein frame such that
\be
\tilde\epsilon = \frac 12 \(\frac 1{U}\frac{\partial U}{\partial \tilde\Phi}\)^2\,,
\label{epsilon-e}
\ee
one can simply show that
\be
\tilde\epsilon = \frac 12 \(\frac{F^2}{V}\frac{\partial \Phi}{\partial\tilde\Phi}\frac{\partial }{\partial \Phi}\(\frac{V}{F^2}\)\)^2
= \epsilon + \ft\,.
\label{eps-e-eps-j}
\ee
It is well known that the power spectrum for the scalar perturbation generated from inflaton field $\tilde\Phi$ in the Einstein frame is given by
\be
{\cal P}_{\zeta} \simeq  \left.\frac{U}{24\pi^2 \tilde\epsilon}\right |_{k |\tau| = 1}\,,
\label{pr-e}
\ee
where the above expression is evaluated at the conformal time $\tau$ when the perturbation with wavenumber $k$ exits the horizon,
and the tensor-to-scalar ratio is
\begin{eqnarray}
r  \simeq16\tilde\epsilon\,.
\label{t2s-e}
\end{eqnarray}
Since the power spectra are frame-independent, we can use Eq.(\ref{eps-e-eps-j}) to write the power spectrum in Eq.(\ref{pr-e})
and the tensor-to-scalar ratio in Eq.(\ref{t2s-e})
in terms of $\epsilon$ as
\ba
{\cal P}_{\zeta} &\simeq& \left.\frac{V}{24\pi^2 F\Big(\epsilon + {\cal F}_{t}\Big)}\right |_{k |\tau| = 1}\,,
\label{pr-cal}\\
r &\simeq&16\left(\epsilon+{\cal F}_{t}\right)\,.
\label{t2s}
\ea
The spectrum index for this power spectrum can be computed via
\be
n_s = \frac{d\ln {\cal P}_{\zeta}}{d \ln k} + 1
\simeq \frac 1{H} \frac{d\ln {\cal P}_{\zeta}}{d t} + 1 \simeq 1 - 2\epsilon - 2 {\cal F}_{t}
- \Phi'\frac{d\ln(\epsilon+{\cal F}_{t})}{d\Phi}\,,
\label{ns}
\ee
Using Eq.(\ref{epsi-ss}), we get
\begin{eqnarray}
n_s &=& 
1 + \frac{1}{\varphi^2 \Big(G_0 \varphi^2 \xi + G_0 + 6 \varphi^2 \xi^2 \Big)^2} \times\nonumber\\&&\Big[2 \Big(\(\varphi^3 \xi +\varphi \)^2 \Theta_{\varphi} \(G_0 \varphi^2 \xi + G_0 + 6 \varphi^2 \xi^2\) - 4 \(G_0 \(2 \varphi^4 \xi^2 + 5 \varphi^2 \xi + 3\) +12 \varphi^2 \xi^2 \(\varphi^2 \xi + 2\)\)\Big)  \nonumber\\&&+ 2 \varphi \Theta \(\varphi^2 \xi +1\) \(G_0 \(\varphi^4 \xi^2 - 3 \varphi^2 \xi -4\) + 6 \varphi^2 \xi^2 \(\varphi^2 \xi - 5\)\)  \nonumber\\&&+ \Theta^2 \(-\(\varphi^3 \xi +\varphi \)^2\) \(G_0 \varphi^2 \xi + G_0 +6 \varphi^2 \xi^2\)\Big]\,,
\label{ns-noexp}
\end{eqnarray}
where $\Theta_\varphi = \partial \Theta / \partial \varphi$.
For our discussion below, we expand the spectrum index around $\xi$ for both $\xi\ll 1$ and $\xi\gg 1$. For the first case $\xi\ll 1$, we find
\begin{eqnarray}
n_s \simeq 1-\frac{\Big(24+8\Theta \varphi+\Theta^{2}\varphi^{2}\Big)}{G_{0}\varphi^{2}} +\frac{1}{G_{0}}\Big[8+2\Theta\varphi-\Theta^{2}\varphi^{2}\Big]\xi+{\cal O}(\xi^2)\,,
\label{ns-sxi}
\end{eqnarray}
and for $\xi\gg 1$
\begin{eqnarray}
n_s &\simeq& 1+\frac{1}{6}\Theta \varphi\Big(2-\Theta\varphi\Big) + \frac{1}{36\xi\varphi^{2}}\Big(-96-48\Theta \varphi - 12\Theta^{2}\varphi^{2}-2\Theta G_{0}\varphi^{3}+\Theta^{2} G_{0}\varphi^{4}\Big) \nonumber\\&&+{\cal O}(1/\xi^2)\,.
\label{ns-lxi}
\end{eqnarray}
The amplitude of the curvature perturbation can be directly read from the power spectrum and we find
\begin{eqnarray}
|\zeta|^2 \simeq \left.\frac{V}{24\pi^{2} F^2\Big(\epsilon + {\cal F}_{t}\Big)}\right |_{c_{s} k|\tau| = 1}\,,
\label{zeta2}
\end{eqnarray}
Using Eq.(\ref{epsi-ss}), the above equation becomes
\be
|\zeta|^2 = \frac{\varphi^2 \(G_0 \varphi^2 \xi + G_0 +6 \varphi^2 \xi^2\) V(\varphi)}
{12 \pi^2 \(\varphi^2 \xi + 1\)^2 \(\Theta \(\varphi^3 \xi + \varphi \) + 4\)^2}\,.
\label{zeta2-noexp}
\ee
The expansion of $|\zeta|^2$ around $\xi$ for both $\xi\ll 1$ and $\xi\gg 1$ respectively reads
\begin{eqnarray}
|\zeta|^2 \simeq \frac{G_{0}\varphi^{6}f(\varphi)}{12\pi^2(4+\Theta \varphi)^{2}} - \frac{G_{0}f(\varphi)}{12\pi^2(4+\Theta \varphi)^{3}}\Big[\varphi^{8}(4+3\Theta \varphi)\Big]\xi+{\cal O}(\xi^2)\,,\label{zeta2-sxi}
\end{eqnarray}
and
\begin{eqnarray}
|\zeta|^2 \simeq \frac{f(\varphi)}{2\pi^{2}\Theta^{2}\varphi^{2}\xi^{2}} + \frac{\varphi^{-5}}{12\pi^{2}\Theta^{3}\xi^{2}}\Big[-48-24\Theta \varphi+G_{0}\Theta \varphi^{3}\Big]f(\varphi)+{\cal O}(1/\xi^4)\,,
\label{zeta2-lxi}
\end{eqnarray}
We can write the tensor-to-scalar ratio in terms of functions $F$ and $G$ as
\be
r = 16\frac{\(\Theta \(\varphi^3 \xi + \varphi \) + 4\)^2}
{2 \varphi^2 \(G_0 \varphi^2 \xi + G_0 + 6 \varphi^2 \xi^2 \)}\,.
\label{t2s-noexp}
\ee
The  $\xi$-expansion of the tensor-to-scalar ratio $r$ reads for $\xi\ll 1$
\begin{eqnarray}
r \simeq \frac{8(4+\Theta \varphi)^{2}}{G_{0}\varphi^{2}} + \frac{8}{G_{0}}\left(-16+\Theta^{2}\varphi^2\right)\xi  +{\cal O}(\xi^2)\,.
\label{t2s-sxi}
\end{eqnarray}
and for $\xi\gg 1$
\begin{eqnarray}
r \simeq \frac{4\Theta^{2}\varphi^2}{3} + \frac{2\left(48\Theta +12\Theta^{2} \varphi - G_{0}\Theta^{2}\varphi^{3}\right)}{9\xi\varphi}+{\cal O}(1/\xi^2)\,.
\label{t2s-lxi}
\end{eqnarray}

In the next section, we will implement our setup to investigating the models of composite inflation.

\section{Contact with Observations}
\label{model}
As it is well known in particle physics, the elementary scalar sector, like the Higgs field in the SM, is plagued by the so-called hierarchy problem. Therefore, a natural hope for employing the composite fields is to solve such problem. In this section, we examine observational constraints on single-field inflation in which the inflaton is a composite field stemming from a four-dimensional strongly interacting field theory non-minimally coupled to gravity using the Planck data
and BICEP2 results. This non-minimal coupling of scalar fields to gravity was pioneered in several earlier \cite{Spokoiny:1984bd,Futamase:1987ua,Salopek:1988qh,Fakir:1990eg,Komatsu:1999mt,Tsujikawa:2004my} and recent works \cite{Bezrukov:2007ep} where a similar phenomenological large value of $\xi$ was needed. However, it is very interesting to further study a potential origin of such a large coupling. It was potentially shown that the models of composite inflation nicely respect tree-level unitary for the scattering of the inflation field during inflation all the way to the Planck scale \cite{Channuie:2011rq,Bezrukov:2011mv,Channuie:2012bv}.

\subsection{Minimal Composite Inflation: $f(\Phi) \equiv \kappa/4$}
The authors of \cite{Channuie:2011rq} recently demonstrated that it is possible to obtain a successful inflation in which the inflaton is a composite field stemming from a four-dimensional strongly interacting field theory. In this work, they engaged the simplest models of technicolor passing precision tests well known as the minimal walking technicolor (MWT) theory \cite{Sannino:2004qp,Hong:2004td,Dietrich:2005wk,Dietrich:2005jn} with the standard (slow-roll) inflationary paradigm as a template for composite inflation. A natural hope is to tie the two mechanisms in such a way that only one natural dynamical mechanism is sufficient to break the electroweak symmetry and also lead inflation. However, within the framework present in \cite{Channuie:2011rq}, the authors showed that the dynamical scale behind inflation is the grand unified energy scale. Therefore the new strongly coupled dynamics cannot be associated directly to electroweak symmetry breaking.
\begin{figure}
\begin{center}
\includegraphics[width=5.5in]{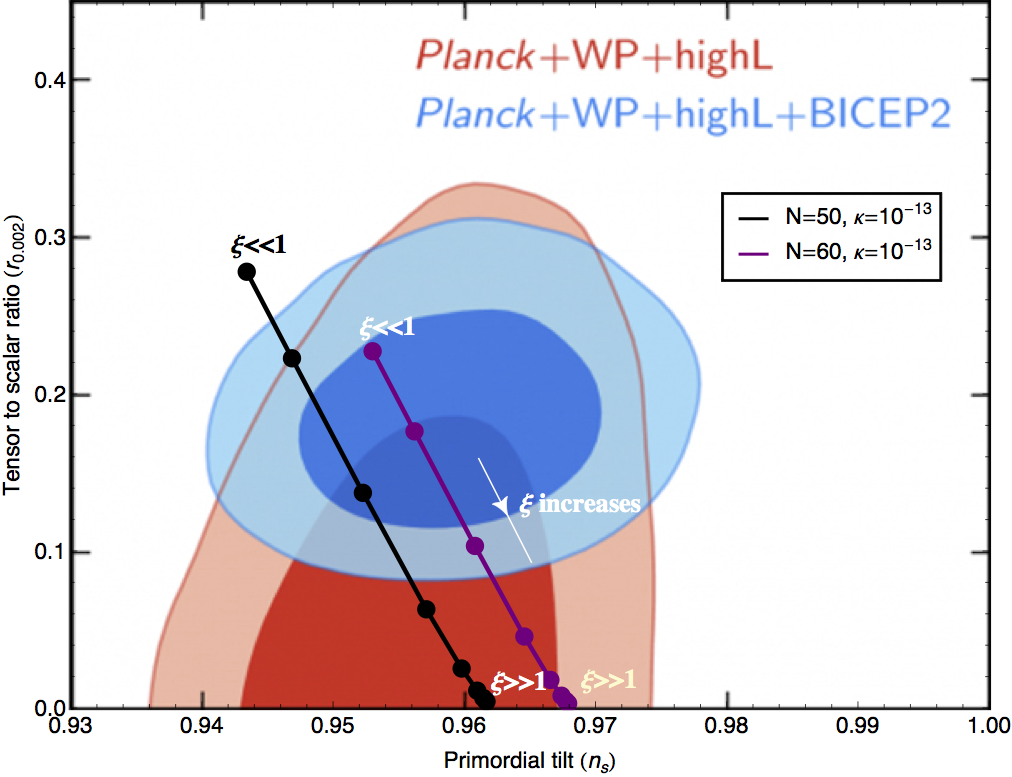} 
\end{center}
\caption{The contours display the resulting 68$\%$ and 95$\%$ confidence regions for the tensor-to-scalar ratio $r$ and the scalar spectral index $n_{s}$. The red contours are for the Planck+WP+highL data combination, which for this model extension gives a 95$\%$ bound $r < 0.26$ (Planck Collaboration XVI 2013 \cite{Planck}). The rest represents the BICEP2 constraints on $r$. The plots show the numerical data resulting from the MCI model by varying the coupling $\xi$ within a range of $10^{-3}\lesssim \xi\lesssim 10^{6}$.} 
\label{f1tc}
\end{figure}
\begin{figure}
\centering
\includegraphics[width=6.0in]{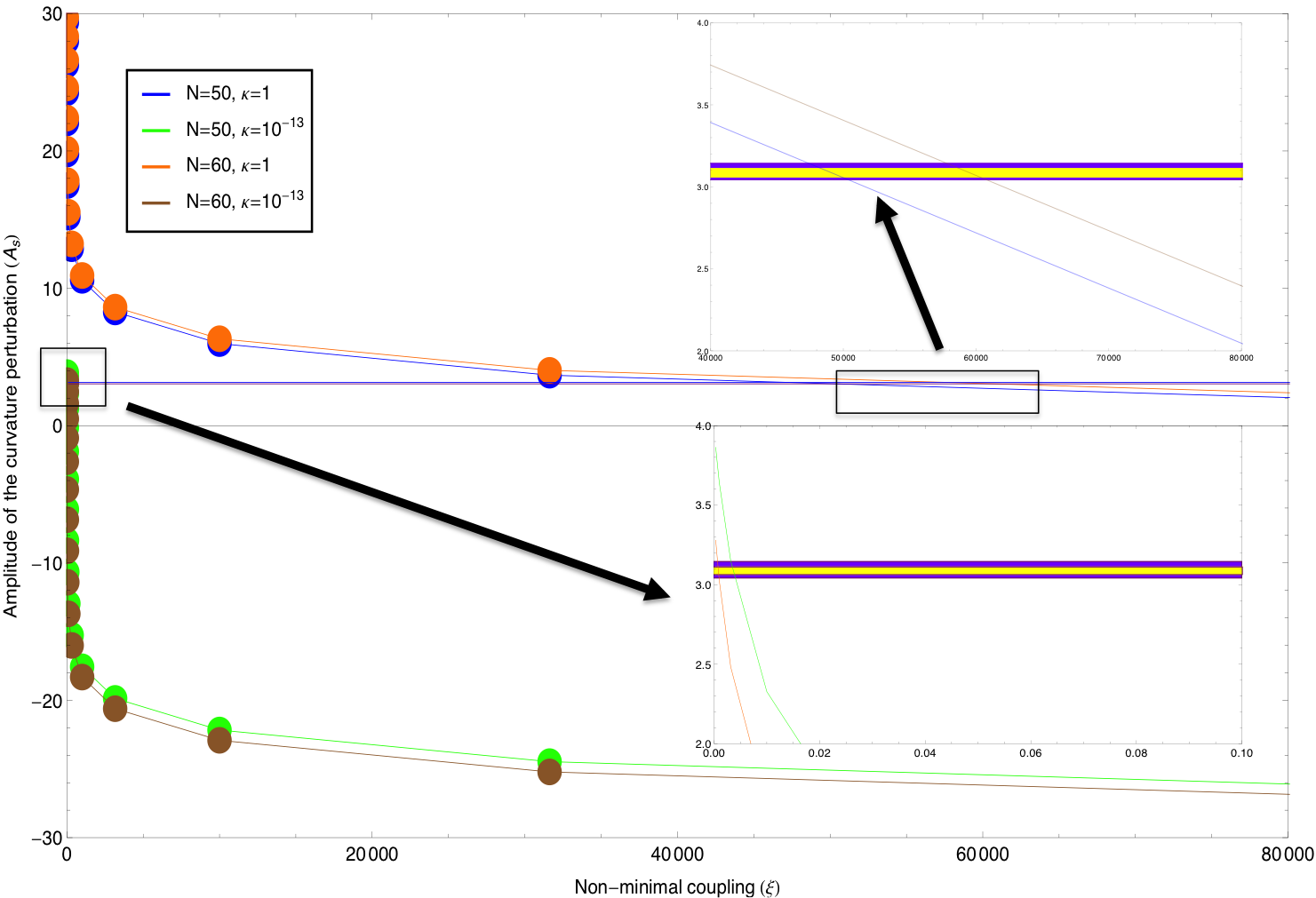}
\caption{The plot shows the relation between the amplitude of the power spectrum ${\cal A}_{s}$ and the non-minimal coupling $\xi$ within a range of $10^{-3}\lesssim \xi\lesssim 10^{6}$ for ${\cal N}=50,60$ predicted by the MCI model. The horizontal bands represent the $1\sigma$ (yellow) and $2\sigma$ (purple) CL for ${\cal A}_{s}$ obtained from Planck. 
}
\label{f1-zeta}
\end{figure}
For model of inflation, the inflaton in this consideration is identified with the lightest composite state, and we write $\Phi\equiv\varphi$, in which we couple non-minimally to gravity. The resulting action in the Jordan frame is given by \cite{Channuie:2011rq}:
\begin{eqnarray}
{\cal S}_{\rm TI}=\int d^{4}x\sqrt{-g}\left[\frac{1 +\xi\varphi^{2}}{2} R - \frac{1}{2}g^{\mu\nu} \partial_{\mu}\varphi\partial_{\nu}\varphi- V_{\rm TI}(\varphi)  \right]\,,
\end{eqnarray}
where
\begin{eqnarray}
V_{\rm TI}(\varphi) = -\frac{m^2}{2}\varphi^2+\frac{\kappa}{4}\varphi^4\,,
\label{v-tc}
\end{eqnarray}
in which $\kappa$ is a self coupling and the inflaton mass is $m^{2}_{\rm TI}=2m^2$.
Since $m_{\rm TI}$ is order of the GeV energy scale, $\kappa$ should be order of unity and $\varphi$ during inflation is order of Planck mass,
we neglect $m^{2}_{\rm TI}$ term in our calculation.
For this model, we have
\begin{eqnarray}
F(\varphi) = 1 + \xi\varphi^{2} \quad {\rm and} \quad G = G_{0}= 1\,.
\label{fg-tc}
\end{eqnarray}
For this form of the potential, Eqs.(\ref{ns-noexp}), (\ref{t2s-noexp}), (\ref{zeta2-noexp}) and (\ref{efold}) yield the intuitive results,
\ba
n_s &\simeq & 1 - \frac{8 \(2 \varphi^4 \xi^2 \(6 \xi + 1\) + \varphi^2 \xi \(24 \xi + 5\) + 3\)}
{\(\varphi^3 \xi \(6 \xi +1\) + \varphi \)^2}\,,
\quad
r \simeq \frac{128}{\varphi^4 \xi \(6 \xi + 1\) + \varphi^2}\,,
\\
|\zeta|^2 &\simeq & \frac{\kappa \varphi^6 \(\varphi^2 \xi \(6 \xi + 1\) + 1\)}
{768 \(\pi \varphi^2 \xi +\pi \)^2}\,,
\quad
{\cal N} \simeq 
\frac{1}{8} \(\(6 \xi +1\)\(\varphi^2 - \varphi_e^2\) - 6 \log\(\frac{\varphi^2 \xi +1}{\varphi_e^2 \xi +1}\) \)\,.
\ea
For more transparency, we consider in the large $\xi$ limit, and write $n_s$, $r$ and $|\zeta|^2$ in terms of ${\cal N}$ as
\ba
n_s &\simeq& 1-\frac{8}{3 \varphi^2 \xi} + {\cal O}(1/\xi^{2})
\simeq 1 - \frac 2{{\cal N}}\,,
\label{ns-tc-l}\\
r &\simeq& \frac{64}{3\varphi^4 \xi^2} - \frac{32}{9\varphi^{4} \xi^3} + {\cal O}(1/\xi^{4})
\simeq \frac{12}{{\cal N}^2}\,,
\label{ns-tc-s}\\
|\zeta|^2 &\simeq& \frac{\kappa \varphi^4}{128\pi^2} +\frac{\kappa (-12\varphi^2 + \varphi^{4})}{768\pi^2 \xi}+ {\cal O}(1/\xi^2) 
\simeq \frac{\kappa{\cal N}^{2}}{72\pi^{2}\xi^{2}}\,.
\label{zmci}
\ea
In this case with $\kappa \sim {\cal O}(1)$, $n_{s},\,r$ and ${\cal A}_{s}$ are well consistent with the Planck data up to $2\sigma$ CL for ${\cal N}=60$ and $4.7\times 10^{4}\lesssim \xi \lesssim 5.0\times 10^{4}$, for instance, illustrated in Fig\,(\ref{f1-zeta}). However, ${\cal A}_{s}$ does strongly depend on ${\cal N}$, and thus the coupling can be lowered (or raised) if ${\cal N}$ changes. Unfortunately, with large $\xi$, the prediction $r$ for this model is in tension with the recent BICEP2 results shown in Fig\,(\ref{f1tc}). In the small $\xi$ limit, we have
\ba
n_s &\simeq& \Big(1-\frac{24}{\varphi^2}\Big)+8\xi + {\cal O}(\xi^2)
\simeq 1 - \frac 3{{\cal N}}\,,
\\
r &\simeq& \frac{128}{\varphi^2} - 128 \xi + {\cal O}(\xi^2)
\simeq \frac{16}{{\cal N}}\,,
\label{t2s-tc-s}\\
|\zeta|^2 &\simeq& \frac{\kappa \varphi^6}{768 \pi ^2}- \frac{\kappa \varphi^8 \xi}{768\pi^2} + {\cal O}(\xi ^2)
\simeq \frac{2 \kappa {\cal N}^3}{3 \pi ^2} \,.
\end{eqnarray}
In this situation, the prediction of $r$ for this model can be consistent with the recent BICEP2 results shown in Fig\,(\ref{f1tc}). In addition, we discover that the amplitude of the curvature perturbation ${\cal A}_{s}$ can satisfy the Planck data at $2\sigma$ CL for ${\cal N}=60,\,\xi \sim 10^{-3}$ if $\kappa\sim 10^{-13}$ illustrated in Fig\,(\ref{f1-zeta}). However, in this case the prediction of such a vary small $\kappa$ is opposed to that from the underlying theory $\kappa \sim {\cal O}(1)$. We will further provide detailed discussions about this model for a small and large $\xi$ in the last section.

\subsection{Glueball Inflation: $f(\Phi)\propto\ln(\Phi/\Lambda^{4})$}
In the current investigation, we consider another viable model of composite inflation. The simplest, but intuitive, examples of strongly coupled theories are pure Yang-Mills theories featuring only gluonic-type fields. The authors of \cite{Bezrukov:2011mv} demonstrated that it is possible to achieve successful inflation where the inflaton emerges as the interpolating field describing the lightest glueball associated to a pure Yang-Mills theory. The original derivation of the low-energy effective Lagrangian can be found in \cite{Schechter:1980ak,Migdal:1982jp,Cornwall:1983zb}. It is worthy to note here that the theory we are using describes the ground state of pure Yang-Mills theory, and of course is not the simple $\phi^4$ theory. Here one can effectively solve the cosmological \lq\lq hierarchy problem\rq\rq\, in the scalar sector of the inflation which is not solved by Higgs inflation. According to the model we are considering, the scalar field corresponds to a low energy effective action which accounts already for the underlying quantum non-perturbative corrections of pure Yang-Mills. Therefore, the model needs not to be further renormalized. 
\begin{figure}
\begin{center}
\includegraphics[width=5.5in]{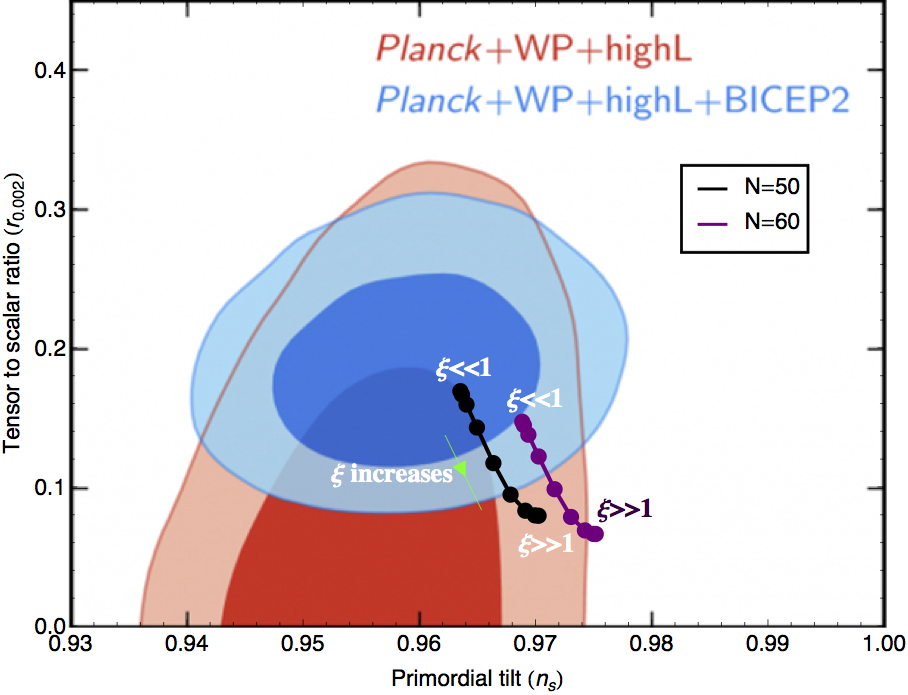} 
\end{center}
\caption{The contours display the resulting 68$\%$ and 95$\%$ confidence regions for the tensor-to-scalar ratio $r$ and the scalar spectral index $n_{s}$. The red contours are for the Planck+WP+highL data combination, which for this model extension gives a 95$\%$ bound $r < 0.26$ (Planck Collaboration XVI 2013 \cite{Planck}). The rest represents the BICEP2 constraints on $r$. The plots show the numerical data resulting from the GI model by varying the coupling $\xi$ within a range of $10^{-3}\lesssim \xi\lesssim 10^{6}$.} 
\label{f1gb}
\end{figure}
\begin{figure}
\centering
\includegraphics[width=6.0in]{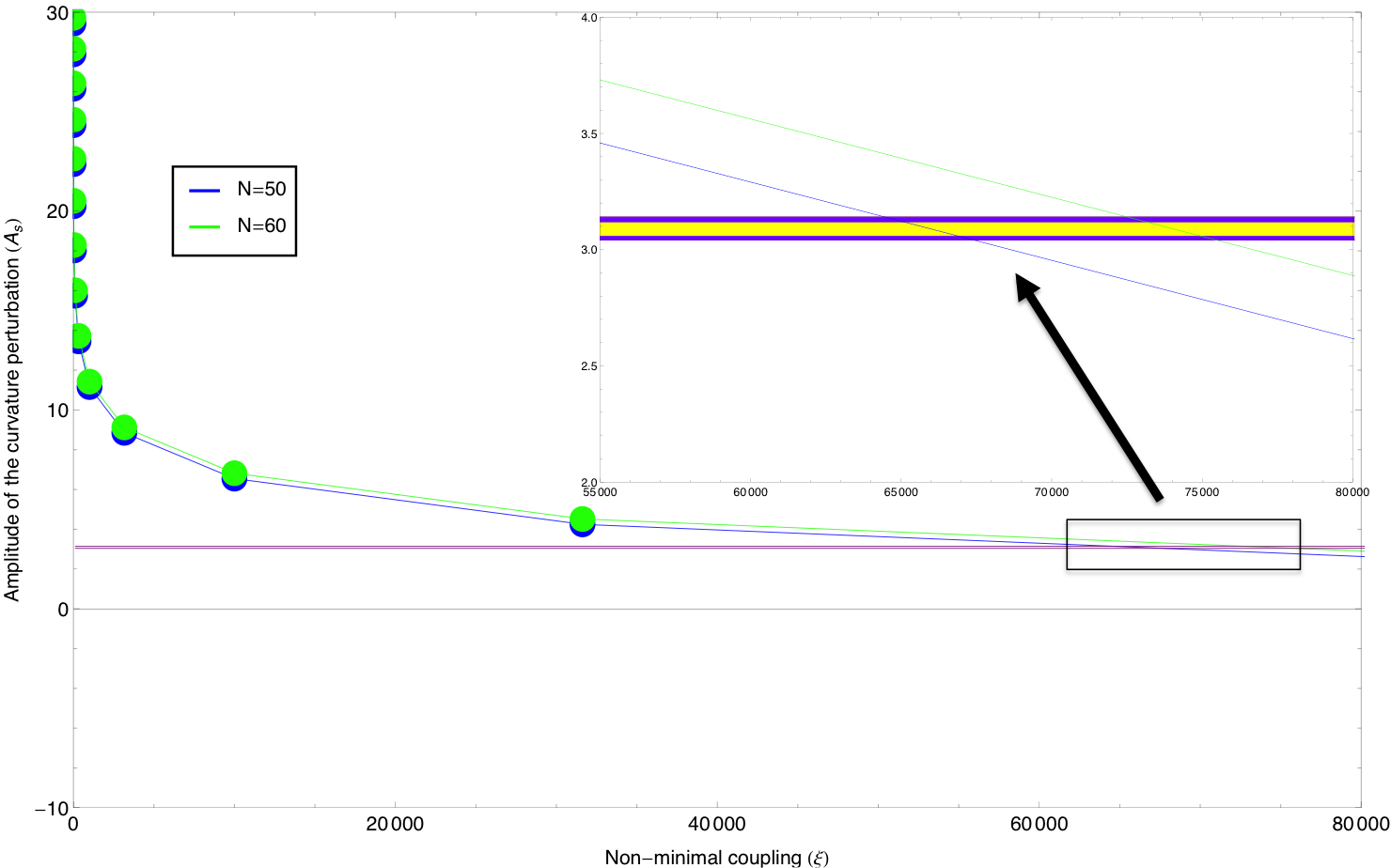}
\caption{The plot shows the relation between the amplitude of the power spectrum ${\cal A}_{s}$ and the non-minimal coupling $\xi$ within a range of $10^{-3}\lesssim \xi\lesssim 10^{6}$ for ${\cal N}=50,60$ predicted by the GI model. The horizontal bands represent the $1\sigma$ (yellow) and $2\sigma$ (purple) CL for ${\cal A}_{s}$ obtained from Planck.
}
\label{gb-zeta}
\end{figure}
Another difference from the $\phi^4$ theory is that the form of the effective potential, before coupling to gravity, is completely fixed by the underlying gauge theory. There are no small couplings in the effective action potential. However, other models of a scalar field non-nimimally coupled to gravity have already been investigated in numerous papers. For instance, the authors in \cite{Barvinsky:2009fy} considered the renormalization group improvement in the theory of the Standard Model Higgs field playing the role of an inflaton with a strong non-minimal coupling to gravity, and in \cite{Barvinsky:2008ia} studied a quantum corrected inflation scenario driven by the Standard Model type potential. For this model, we write
\begin{eqnarray}
 f(\Phi)=\frac{1}{2}\ln(\Phi/\Lambda^{4})\,,
 \end{eqnarray}
so that the effective Lagrangian for the lightest glueball state, constrained by the Yang-Mills trace anomaly, non-minimally coupled to gravity in the Jordan frame reads
\begin{eqnarray}
{\cal S}_{\rm GI} = \int d^{4}x\sqrt{-g}\Big[\frac{1 + \xi\Phi^{1/2}}{2} R - \frac{1}{2}\Phi^{-3/2}g^{\mu\nu} \partial_{\mu}\Phi\partial_{\nu}\Phi - V_{\rm GI}(\Phi)\Big]\,,
\end{eqnarray}
where
\begin{eqnarray}
V_{\rm GI}(\Phi) = \frac{\Phi}{2}\ln\left(\frac{\Phi}{\Lambda^{4}}\right)\,,
\label{v-gb}
\end{eqnarray}
Here the physical meaning of $\Lambda$ underlying inflationary scenario has already been discussed in \cite{Bezrukov:2011mv}. The effective potential given above is known in particle physics. It is the generating function for the trace anomaly of a generic purely gluonic Yang-Mills theory such that $\Phi$ is a composite operator. It is convenient to introduce the field $\varphi$ possessing unity canonical dimension and related to $\Phi$ as follows: 
\begin{eqnarray}
\Phi=\varphi^{4}\,.
\end{eqnarray}
From the above assignment, the action then becomes
\begin{eqnarray}
{\cal S}_{\rm GI} = \int d^{4}x\sqrt{-g}\Big[\frac{1 + \xi\varphi^{2}}{2} R - \frac{1}{2}g^{\mu\nu} \partial_{\mu}\varphi\partial_{\nu}\varphi - V_{\rm GI}(\varphi)  \Big]\,, \label{varaction}
\end{eqnarray}
where
\begin{eqnarray}
V_{\rm GI}(\varphi) = 2\varphi^{4}\ln\left(\frac{\varphi}{\Lambda}\right)\,,
\label{v-gbr}
\end{eqnarray}
which yields
\begin{eqnarray}
F(\varphi) = 1 + \xi\,\varphi^{2} \quad {\rm and} \quad G_{0} = 16\,.
\label{fg-gbr}
\end{eqnarray}
Here the above quantities can be directly read from the action (\ref{varaction}).
For this form of the potential, $F$ and $G$, we can use Eq.(\ref{efold}) to compute the number of e-foldings in the $\xi \gg 1$ and $\xi \ll 1$ limits respectively as
\ba
{\cal N} \simeq 3\Big(\ln^2\(\varphi/\Lambda\) - \ln^2\(\varphi_{e}/\Lambda\)\Big) + {\cal O}(1/\xi)\,,
\ea
and
\ba
{\cal N} \simeq 
\int_{\varphi_e}^\varphi \frac{16 \tilde\varphi \ln\(\tilde\varphi / \Lambda\)d\tilde\varphi}{4 \ln\(\tilde\varphi/\Lambda\) + 1} + {\cal O}(\xi)
\simeq 2\varphi^2 - 2 \varphi_e^2 + {\cal O}\(\xi\)\,,
\ea
where in the calculation of ${\cal N}$ for the small $\xi$ case, we have supposed that $\ln(\varphi / \Lambda) \gg 1$. This is so since when $\ln(\varphi / \Lambda) < 1$ we get ${\cal N} < 0$ unless $\varphi < \varphi_e$.
However, this approximation for ${\cal N}$ at small $\xi$ limit is valid if $\Lambda \lesssim 1$,
since the ${\cal O}(\xi)$ term in Eq.(\ref{pp-sxi}) becomes significant when $\Lambda > 1$.
In contrary, for $\xi\gg 1$ case, we compute the expression for ${\cal N}$ by supposing that $\Lambda$ is not too small,
otherwise we have to keep the next leading order terms in the expansion of $\varphi'$ and $\epsilon$ given in Eqs.\,(\ref{ep-lxi}) and (\ref{pp-lxi}).
In the case where $\Lambda$ is not too small, it follows from Eq.(\ref{ep-lxi}) that
$\ln\(\varphi_e/\Lambda\) \sim {\cal O}(1)$ at leading order.
Moreover, we see from Eq.(\ref{ep-sxi}) that $\varphi_e \sim {\cal O}(1)$ when $\Lambda < 1$.
Hence, for these cases, we neglect the $\varphi_e$-dependent terms in the above expressionf for ${\cal N}$.
Using these expression for ${\cal N}$, we can write $\varphi$ in terms of ${\cal N}$.
Therefore we can use Eqs.(\ref{ns-lxi}), (\ref{zeta2-lxi}) and (\ref{t2s-lxi}) to write $n_s$, $r$ and $|\zeta|^2$ in terms of ${\cal N}$ for $\xi \gg 1$ case as
\begin{eqnarray}
n_s &\simeq& 1 - \frac{1}{2\ln^2(\varphi / \Lambda)}  + {\cal O}(1/\xi)
\simeq 1-\frac{3}{2{\cal N}} + {\cal O}(\xi)\,,
\label{ns-gb-l}\\
r &\simeq& \frac{4}{3\ln^2(\varphi / \Lambda)}  + {\cal O}(1/\xi)
\simeq \frac{4}{{\cal N}} + {\cal O}(\xi)\,,
\label{t2s-gb-l}\\
|\zeta|^2 &\simeq& \frac{\ln^3\left(\varphi/\Lambda\right)}{\pi^{2}\xi^2}  + {\cal O}(1/\xi^3)
\simeq \frac{{\cal N}^{3/2}}{3\sqrt{3}\pi^2\xi^2} + {\cal O}(1/\xi^3)\,,
\label{zeta2-gb-l}
\end{eqnarray}
and use Eqs.(\ref{ns-sxi}), (\ref{zeta2-sxi}) and (\ref{t2s-sxi}) to express $n_s$, $r$ and $|\zeta|^2$ in terms of ${\cal N}$ for $\xi \ll 1$ case as
\begin{eqnarray}
n_s &\simeq& 1 - \frac{3}{2\varphi^2} - \frac{3}{16\varphi^{2}\ln^2\left(\varphi / \Lambda\right)} - \frac{5}{4\varphi^{2}\ln\left(\varphi/\Lambda\right)} + {\cal O}(\xi)
\simeq  1 - \frac 3{{\cal N}} + {\cal O}(\xi)\,,
\label{ns-gb-s}\\
r &\simeq& \frac{8}{\varphi^{2}} + \frac{1}{2\varphi^{2}\ln^2\left(\varphi/\Lambda\right)} + \frac{4}{\varphi^{2}\ln\left(\varphi/\Lambda\right)}+ {\cal O}(\xi)
\simeq \frac{16}{{\cal N}} + {\cal O}(\xi)\,,
\label{t2s-gb-s}\\
|\zeta|^2 &\simeq& \frac{8\varphi^{6}\ln^3(\varphi/\Lambda)}{3\pi^{2}(1+4\ln(\varphi/\Lambda))^{2}} + {\cal O}(\xi)
\simeq \frac{{\cal N}^3}{48\pi^2} + {\cal O}(\xi)\,.
\label{zeta2-gb-s}
\end{eqnarray}
From the above analytical estimations, we see that when $\xi \gg 1$,
$n_s$ , $r$ and $|\zeta|^2$ can satisfy the 95$\%$CL observational bound from Planck data if $50 < {\cal N} < 60$ and $\xi \sim 10^{4}$. Nevertheless, for such range of ${\cal N}$, $r$ lies outside the $2\sigma$ CL with BICEP2 results shown in Fig.\,(\ref{f1gb}).
The value of $r$ will increase and then satisfy the bound from BICEP2 results when ${\cal N}\lesssim 45$ or $\xi \ll 1$. For the $\xi \ll 1$ limit, however, it follows from Eq.(\ref{zeta2-gb-s}) that $|\zeta|^2$ is much larger than the observational bound from Planck data.
We see from Fig.\,(\ref{f1gb}) and (\ref{gb-zeta}) that
these analytical estimations are in good agreement with the values for $n_s$, $r$ and $|\zeta|^2$ numerically computed from Eqs.(\ref{ns-noexp}), (\ref{t2s-noexp}) and (\ref{zeta2-noexp}).
Implementing numerical analysis allows us to explore the behaviour of $n_{s},\,r$ and $|\zeta|^{2}$ to what extent they are responsible for the confining scale $\Lambda$. From numerical investigations, we come up with the following additional information:
Firstly, for large $\xi$ case, $n_s$ and $|\zeta|^2$ get higher while $r$ gets smaller when $\Lambda$ decreases.
Secondly, $n_s$ and $r$ get higher when $\Lambda$ increases above unity for $\xi \ll 1$ case.

In general, the potential arises in this model of composite inflation is quite subtle,
because it becomes negative when $\varphi < \Lambda$ and its minimum is also negative. However, the authors of \cite{Felder:2002jk} investigated cosmological evolution in models in which the effective potential become negative at some values of the inflaton field. They also discovered several qualitatively new features as compared to those of the positive one. In the next section, we will consider the another compelling paradigm for composite inflation model
that leads to a new form of the potential for inflation.

\subsection{Super-glueball Inflation: $f(\Phi)\propto\ln^{2}(\Phi/\Lambda^{3})$}
This model has been explored in \cite{Channuie:2012bv} in the context of four-dimensional strongly interacting field theories non-minimally coupled to gravity. The authors showed that it is viable to achieve successful inflation driven by orientifold field theories. When the number of colors $N_c$ is large, such theories feature super Yang-Mills properties. In this investigation, we assign the inflaton as the gluino-ball state in SYM theory. The effective Lagrangian in supersymmetric gluodynamics was constructed in 1982 by Veneziano and Yankielowicz (VY) \cite{GVSY}. The component bosonic form of the VY Lagrangian was summarized in \cite{Sannino:2003xe}. For this model, we write
\begin{eqnarray}
 f(\Phi)=\frac{4\alpha N^{2}_{c}}{9}\ln^{2}(\Phi/\Lambda^{3})\,.
 \end{eqnarray}
As always investigated in standard fashion, we take the scalar component part of the super-glueball action and coupled it non-minimally to gravity. Focusing only on the modulus of the inflaton field and taking the next step in order to write the non-minimally coupled scalar component part of the super-glueball action to gravity, the resulting action in the Jordan frame reads
\begin{eqnarray}
{\cal S}_{\rm SgbI} = \int d^{4}x\sqrt{-g}\Big[\frac{1 + N^{2}_{c}\xi\Phi^{2/3}}{2} R - \frac{N^{2}_{c}}{2\alpha}\Phi^{-4/3}g^{\mu\nu} \partial_{\mu}\Phi\partial_{\nu}\Phi - V_{\rm SgbI}(\Phi)\Big]\,,
\end{eqnarray}
where
\begin{eqnarray}
V_{\rm SgbI}(\Phi) = \frac{4\alpha N^{2}_{c}}{9}\Phi^{4/3}\ln^{2}(\Phi/\Lambda^{3})\,.
\label{v-sgb}
\end{eqnarray}
\begin{figure}
\begin{center}
\includegraphics[width=5.5in]{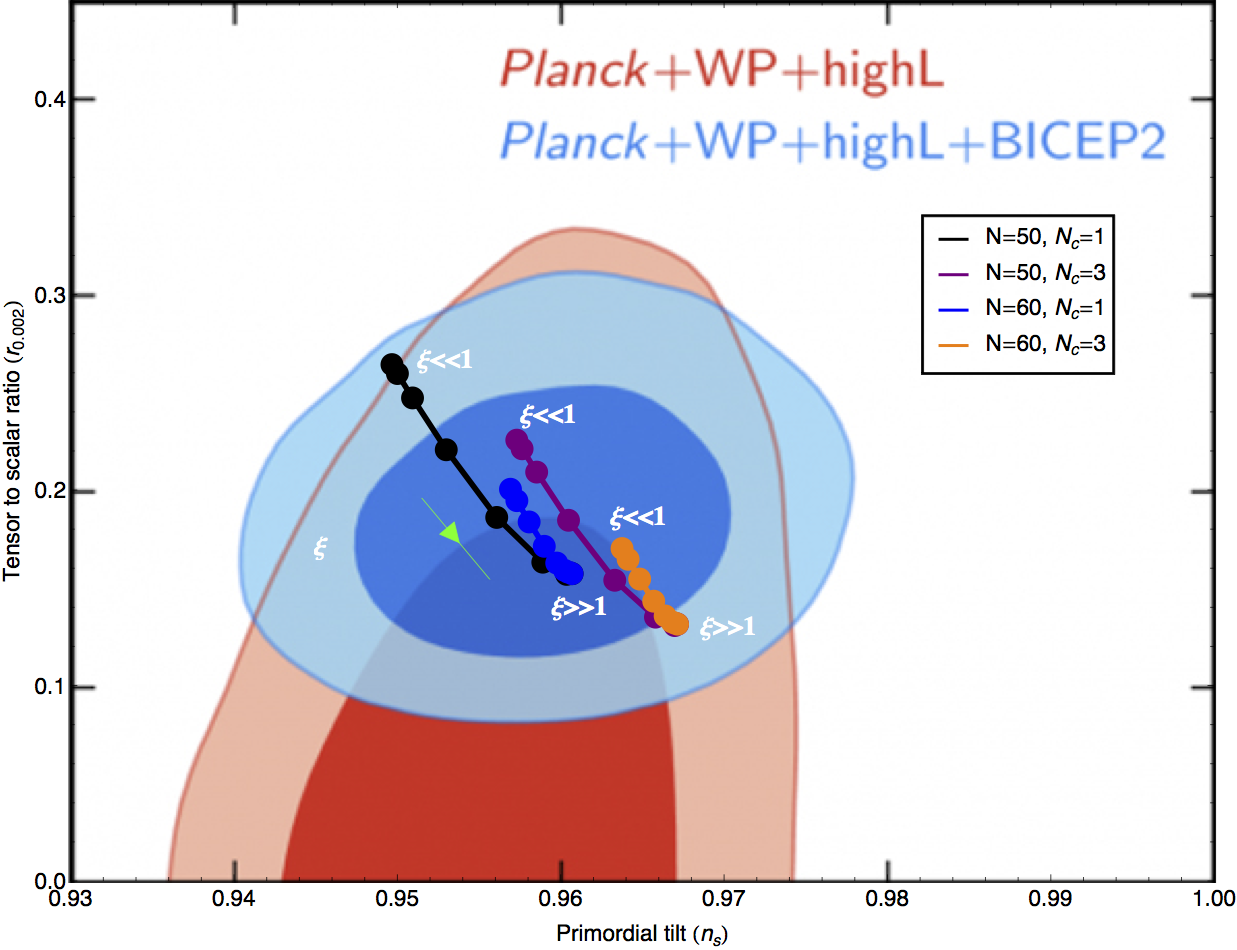} 
\end{center}
\caption{The contours display the resulting 68$\%$ and 95$\%$ confidence regions for the tensor-to-scalar ratio $r$ and the scalar spectral index $n_{s}$. The red contours are for the Planck+WP+highL data combination, which for this model extension gives a 95$\%$ bound $r < 0.26$ (Planck Collaboration XVI 2013 \cite{Planck}). The rest represents the BICEP2 constraints on $r$. The plots show the numerical data resulting from the SgbI model by varying the coupling $\xi$.} 
\label{f1sgb}
\end{figure}
\begin{figure}
\centering
\includegraphics[width=6.0in]{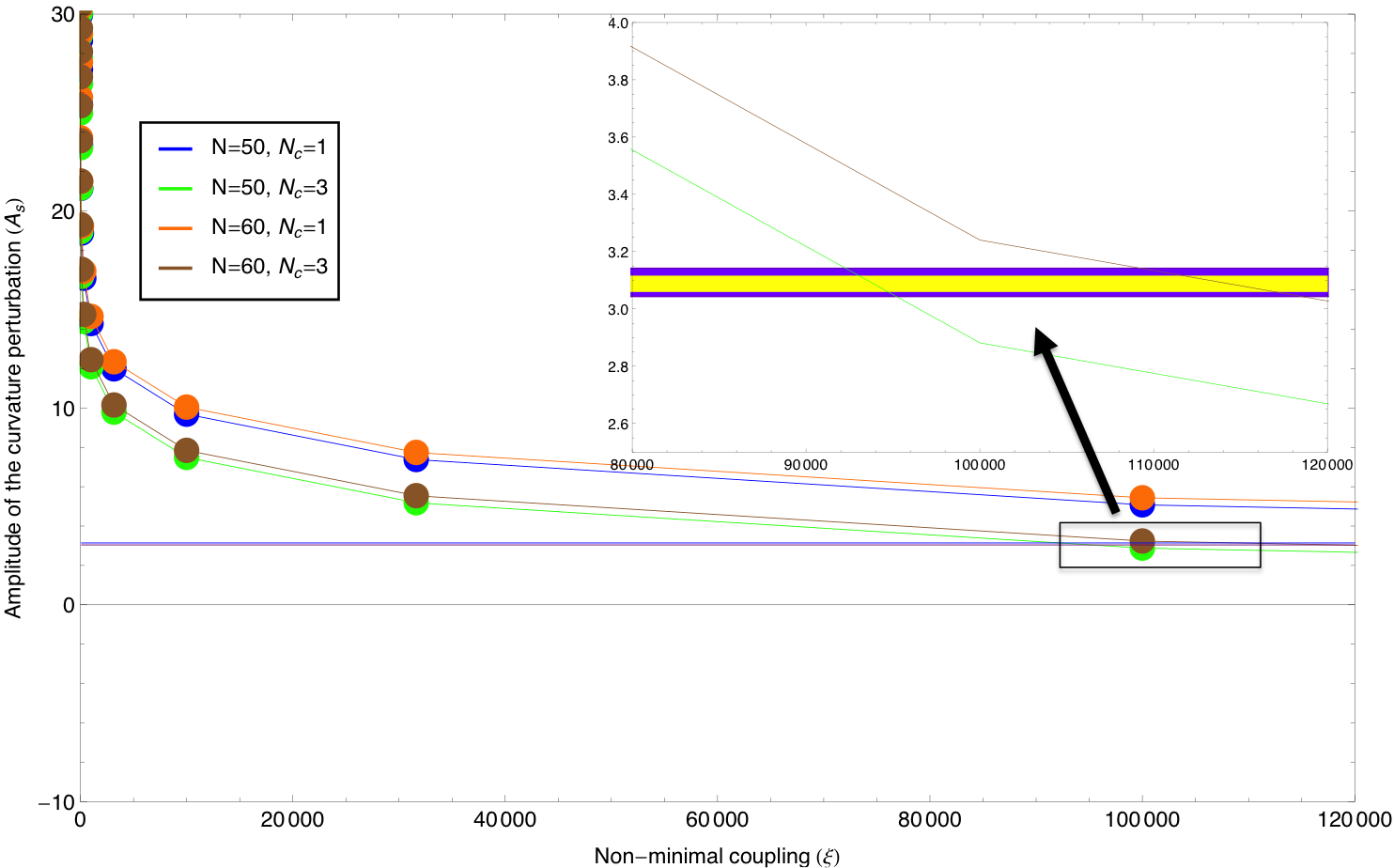}
\caption{The plot shows the relation between the amplitude of the power spectrum ${\cal A}_{s}$ and the non-minimal coupling $\xi$ for ${\cal N}=50,60$ predicted by the SgbI model. The horizontal bands represent the $1\sigma$ (yellow) and $2\sigma$ (purple) CL for ${\cal A}_{s}$ obtained from Planck.
}
\label{f1-sgb}
\end{figure}
In order to re-write the action in terms of the new field $\varphi$ possessing unity canonical dimension, we replace the field $\Phi$ with $\Phi=\varphi^{3}$ since in this case $D=3$. Therefore, the action of the theory for our investigation is given by
\begin{eqnarray}
{\cal S}_{\rm SgbI} = \int d^{4}x\sqrt{-g}\Big[\frac{1+N^{2}_{c}\xi\varphi^{2}}{2} R - \frac{9N^2_{c}}{2\alpha}g^{\mu\nu} \partial_{\mu}\varphi\partial_{\nu}\varphi - V_{\rm SbgI}(\varphi)  \Big]\,,
\end{eqnarray}
where
\begin{eqnarray}
V_{\rm SgbI}(\varphi) = 4\alpha N^{2}_{c}\varphi^{4}(\ln[\varphi/\Lambda])^{2}\,,
\end{eqnarray}
with $N_{c}$ a number of colours. With the action given above, we find
\begin{eqnarray}
F(\varphi) = 1 + N^{2}_{c}\xi\,\varphi^{2} \quad {\rm and} \quad G = \frac{9N^{2}_{c}}{\alpha}\,.
\label{fg-sgb}
\end{eqnarray}
Using the similar approximations to those of the Glueball Inflation,
the number of e-foldings for this inflation model in the large and small $\xi$ limits are respectively approximated by
\ba
{\cal N} \simeq \frac{3}{2}\Big(\ln^2\(\varphi/\Lambda\) - \ln^2\(\varphi_{e}/\Lambda\)\Big) + {\cal O}(1/\xi)\,,
\label{efold-sgb-l}
\ea
and
\ba
{\cal N} \simeq \int_{\varphi_e}^{\varphi}
\frac{9 N_c^2 \varphi \ln\(\varphi/\Lambda\)}{4 \alpha \ln\(\varphi /\Lambda\) + 2 \alpha} + {\cal O}\(\xi\)
\simeq  \frac{9 N_c^2}{8 \alpha }\(\varphi^2 - \varphi_e^2\) + {\cal O}\(\xi\)\,.
\label{efold-sgb-s}
\ea
Regarding to the above relations between the number of e-foldings and $\varphi$,
we can compute $n_s$, $r$ and $|\zeta|^2$ in terms of ${\cal N}$ for $\xi \gg 1$ and $\xi \ll 1$ limits.
For a large $\xi$ limit, Eqs.(\ref{ns-lxi}), (\ref{t2s-lxi}) and (\ref{zeta2-lxi}) yield
\begin{eqnarray}
n_{s} 
&\simeq& 1 - \frac{4}{3\ln^2(\varphi/\Lambda)} + {\cal O}(1/\xi)
\simeq 1 - \frac 2{{\cal N}} + {\cal O}(1/\xi)\,,
\label{ns-sgb-l}\\
r 
&\simeq& \frac{16}{3\ln^2(\varphi/\Lambda)}  + {\cal O}(1/\xi)
\simeq \frac 8{{\cal N}} + {\cal O}(1/\xi)\,,
\label{t2s-sgb-l}\\
|\zeta|^2 &\simeq& \frac{4\alpha\ln^4(\varphi/\Lambda)}{72N^{2}_{c}\pi^{2}\xi^{2}} + {\cal O}(1/\xi^{3})
\simeq \frac{2\alpha{\cal N}^{2}}{81N^{2}_{c}\pi^{2}\xi^{2}} + {\cal O}(1/\xi^3)\,,
\label{zeta2-sgb-l}
\end{eqnarray}
and for a small $\xi$ limit, Eqs.(\ref{ns-sxi}), (\ref{t2s-sxi}) and (\ref{zeta2-sxi}) give
\begin{eqnarray}
n_{s} &\simeq& 1- \frac{24\alpha}{9N^{2}_{c}\varphi^{2}} - \frac{8\alpha}{9N^{2}_{c}\varphi^{2}\ln(\varphi/\Lambda)} - \frac{20\alpha}{9N^{2}_{c}\varphi^{2}\ln^2(\varphi/\Lambda)} +{\cal O}(\xi)
\simeq 1 - \frac 3{{\cal N}} + {\cal O}(\xi)\,,
\label{ns-sgb-s}\\
r &\simeq& \frac{32\alpha\Big(1+4\ln(\varphi/\Lambda) + 4\ln^2(\varphi/\Lambda)\Big)}{9N^{2}_{c}\varphi^{2}\ln^2(\varphi/\Lambda)} + {\cal O}(1/\xi)
\simeq \frac{16}{{\cal N}} + {\cal O}(\xi)\,,
\label{t2s-sgb-s}\\
|\zeta|^2 &\simeq& \frac{N^{4}_{c}\varphi^{6}\ln^4(\varphi/\Lambda)}{12\pi^{2}\Big(1+2\ln(\varphi/\Lambda)\Big)^{2}}+{\cal O}(\xi)
\simeq \frac{64 N_{c}\alpha^3 {\cal N}^3 }{2187 \pi^{2}} + {\cal O}(\xi)\,.
\label{zeta2-sgb-s}
\end{eqnarray}
The main results from the above analytical estimations are similar to those of Glueball Inflation.
The interesting different result is that for this model of inflation, $r$ can be large enough to satisfy the bound launched by BICEP2. We compute $n_s$, $r$ and $|\zeta|^2$ for this model numerically using Eqs.(\ref{ns-noexp}), (\ref{t2s-noexp}) and (\ref{zeta2-noexp}) for a given number of e-foldings, and plot the relevant results in Figs.\,(\ref{f1sgb}) and (\ref{f1-sgb}). Concretely, the predictions of this model are satisfactorily consistent with the BICEP2 data at $68\%$ CL if ${\cal N} \subseteq [50,\,60]$ and $\xi \sim 10^{4}-10^{5}$ with an exceptional range of $\Lambda$. Regarding to such ranges of ${\cal N},\,\xi$ and $\Lambda$, we discover that the predictions of this model can also satisfy the Planck data. Except for the results illustrated in Figs.\,(\ref{f1sgb}) and (\ref{f1-sgb}), the numerical implementations also address that, for $\xi \gg 1$, $r$ gets smaller when $\Lambda$ decreases,
and $r$ is in tension with the BICEP2 results when $\Lambda < 10^{-4}$.

\section{Conclusions}
\label{concl}

In this work, we constrain the model parameters of various composite inflationary models using the observational bound for $n_s$ and $r$ from Planck and recent BICEP2 observations, and use ${\cal A}_{s}$ from Planck data. The general action for the composite inflation has to be in the form of scalar-tensor theory in which the inflaton is non-minimally coupled to gravity. We compute the power spectra for the curvature perturbations by using the usual slow-roll approximations. We summarise our findings, illustrated in Fig.\,(\ref{three}) as follows:
\begin{figure}
\begin{center}
\includegraphics[width=5.5in]{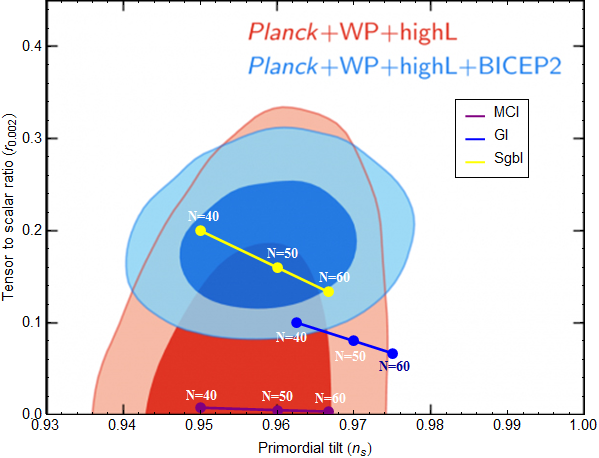} 
\end{center}
\caption{The contours display the resulting 68$\%$ and 95$\%$ confidence regions for the tensor-to-scalar ratio $r$ and the scalar spectral index $n_{s}$. The red contours are for the Planck+WP+highL data combination, which for this model extension gives a 95$\%$ bound $r < 0.26$ (Planck Collaboration XVI 2013 \cite{Planck}). The rest represents the BICEP2 constraints on $r$. The plots summarise the predictions of composite models examined in this work (MCI,\,GI,\,SgbI) assuming the number of e-foldings ${\cal N}$ to the end of inflation lies in the interval [40,\,60] and $\xi\gg1$.} \label{three}
\end{figure}

We discover for MCI model with $\xi\gg 1$ that the predictions lie well inside the joint $68\%$ CL for the Planck+WP+highL data for ${\cal N} = [40,\,60]$, whilst for ${\cal N} = 60$ this model lies on the boundary of $1\sigma$ region of the Planck+WP+highL data. However, with $\xi\gg 1$, the model predictions is in tension with the recent BICEP2 contours. This is so since the model predictions yield quite small values of $r$. Concretely, the model predicts $\epsilon\sim 1/{\cal N}^{2}$ which no longer holds in light of the BICEPS results for $r=16\epsilon$ such that $r=0.2^{+0.07}_{-0.05}$. Nevertheless, this tension can be relaxed if $\xi$ is very small, i.e. $\xi \sim 10^{-3}$. If this is the case, ${\cal A}_{s}$ cannot satisfy the Planck data unless $\kappa$ gets extremely small, i.e. $\kappa\sim 10^{-13}$. Unfortunately, the prediction with very small $\kappa$ is opposed to the underlying theory. This model predicts $n_{s}\simeq 0.960$ and $r\simeq 0.0048$ for ${\cal N}=50$ with $\xi\gg 1$. Likewise, the Higgs inflation is also in tension with the recent BICEP2 data. The authors in \cite{Cook:2014dga} claim the incompatibility between the BICEP2 results and the predictions of Higgs inflation. This tension can be alleviated with the presence of sizable quantum departures from the $\phi^{4}$-Inflationary model with the non-minimally coupled scenario \cite{Joergensen:2014rya}.

The inflationary observables predicted by the GI model lie well inside the Planck+WP+highL data for ${\cal N} \subseteq [45,\,60]$ at the $2\sigma$ region of the contours for large $\xi$. Nevertheless, the predictions of this model with ${\cal N} > 45$ are in tension with the BICEP2 data. However, this can be accommodated to the data allowing that the smaller number of ${\cal N}$ is basically required. It is obvious that ${\cal N}$ is model-dependent quantity. However, it is quite subtle if we have ${\cal N}\lesssim 45$ for model of inflation. This is so since, in order to solve the horizon problem, in the common formulation one frequently use at least ${\cal N}\subseteq [50,\,60]$. We anticipate this can be further verified by studying the reheating effect. 

We discover that ${\cal A}_{s}$ is well consistent with the Planck data up to $2\sigma$ CL for ${\cal N}=60$ with $7.3\times 10^{4}\lesssim \xi \lesssim 7.5\times 10^{4}$, illustrated in Fig\,(\ref{gb-zeta}). However, ${\cal A}_{s}$ does strongly depend on the number of e-foldings implying that the coupling can be lowered (or raised) with changing ${\cal N}$. This model provides $n_{s}\simeq 0.967$ and $r\simeq 0.089$ for ${\cal N}=45$ with $\xi\gg 1$. In the small $\xi$ limit, we find that the predictions for $n_{s}$ and $r$ are consistent with Planck data illustrated in Fig.\,(\ref{f1gb}). In the case $\xi\ll1$, we find that the predictions can also satisfy the recent BICEP2 results. However, with a very small $\xi$ limit, the amplitude of the curvature perturbation ${\cal A}_{s}$ lie far away from the $95\%$ CL of the Planck data shown in Fig.\,(\ref{gb-zeta}).

Surprisingly, the SgbI predictions are fully consistent with BICEP2 constraints for ${\cal N} \subseteq [50, 60]$. Moreover, the model can also be consistent with the Planck contours at $1\sigma$ CL. We discover that ${\cal A}_{s}$ is well consistent with the Planck data up to $2\sigma$ CL for ${\cal N}=50$ and $N_{c}=3$ with $9.2\times 10^{4}\lesssim \xi \lesssim 9.5\times 10^{4}$, illustrated in Fig\,(\ref{f1-sgb}). This model provides $n_{s}\simeq 0.960$ and $r\simeq 0.16$ for ${\cal N}=50$ with $\xi\gg 1$. However, with the very small $\xi$, the amplitude of the curvature perturbation ${\cal A}_{s}$ cannot get close to the $95\%$ CL of the Planck data shown in Fig.\,(\ref{gb-zeta}). It would be nice to use the BICEP2 results to constrain $\Lambda_{\rm SgbI}$ since the data provides us the lower bound on $r$. According to the recent BICEP2 data, we roughly opt $r\simeq 0.12$ and use $N_{c}=1(3)$ predicting $\Lambda_{\rm SgbI}> 10^{-3}(10^{-4})$ which corresponds to, at least, the GUT energy scale in this investigation, in order to satisfy the BICEP2 data at $1\sigma$ CL. We hope that the future observations will provide significant confirmation for this model. 

Another crucial consequence for the model of inflation is the (p)reheating mechanism. We anticipate to investigate this mechanism by following closely references \cite{Bezrukov:2008ut,GarciaBellido:2008ab,Watanabe:2006ku}.

\vspace{0.2in}
\noindent \underline{Acknowledgment}~:~P.C. is fully supported by the \lq\lq Research Fund for DPST Graduate with First Placement\rq\rq\,under Grant No.\,033/2557.~K.K. is supported by Thailand Research Fund (TRF) through grant RSA5480009.

\bibliographystyle{unsrt}

\end{document}